\documentclass{aa} 
\usepackage{graphicx}
\usepackage{color}
\usepackage{graphicx}
\usepackage{lscape}
\usepackage{natbib}
\usepackage[varg]{txfonts}
\usepackage{amssymb}
\usepackage{stmaryrd}
\usepackage{url}
\usepackage{xspace}
\usepackage{color}
\usepackage{siunitx}
\usepackage{textcomp}
\usepackage[usenames,dvipsnames]{xcolor}
\usepackage{hyperref}
\usepackage{longtable}
\usepackage{multicol}
\bibpunct{(}{)}{;}{a}{}{,}

\newcounter{Rco}
\newcommand{\Ionst}[1]{\setcounter{Rco}{#1}\Roman{Rco}}
\newcommand{\Ion}[2]{\mbox{#1\,{\scriptsize\Ionst{#2}}}}
\newcommand{\Ionw}[3]{\mbox{#1\,{\scriptsize\Ionst{#2}}~$\lambda\,#3$\,\AA}}

\newcommand{\Ionww}[3]{\mbox{#1\,{\scriptsize\Ionst{#2}}~$\lambda\lambda\,#3$\,\AA}}

\newcommand{\se}[1]{\mbox{Sect.\,\ref{#1}}}
\newcommand{\logg}{\mbox{$\log g$}\xspace}

\newcommand{\Teff}{\mbox{$T_\mathrm{eff}$}\xspace}

\newcommand{\Msol}{$\mathrm{M}_\odot$}
\newcommand{\Rsol}{$\mathrm{R}_\odot$}

\def\Gaia{{\it Gaia}}

\begin{document} 

\title{Spectroscopic survey of faint planetary-nebula nuclei\\ VI\null. Seventeen hydrogen-rich central stars
\thanks{Based on observations obtained with the Hobby-Eberly Telescope (HET), which is a joint project of the University of Texas at Austin, the Pennsylvania State University, Ludwig-Maximillians-Universit\"at M\"unchen, and Georg-August Universit\"at G\"ottingen. The HET is named in honor of its principal benefactors, William P. Hobby and Robert E. Eberly.}}
\titlerunning{Hydrogen-Rich Planetary Nuclei}

\author{Nicole Reindl \inst{1}
    \and Howard E. Bond \inst{2, 3}
    \and Klaus Werner \inst{4}
    \and Gregory R. Zeimann \inst{5}
    }
    
\offprints{Nicole\,Reindl\\ \email{nreindl885@gmail.com}}

\institute{Landessternwarte Heidelberg, Zentrum f\"ur Astronomie, Ruprecht-Karls-Universit\"at, K\"onigstuhl 12, 69117 Heidelberg, Germany
\and Department of Astronomy and Astrophysics, Penn State University, University Park, PA 16802, USA
\and Space Telescope Science Institute, 3700 San Martin Dr., Baltimore, MD 21218, USA
\and Institut f\"ur Astronomie und Astrophysik, Kepler Center for Astro and Particle Physics, Eberhard Karls Universit\"at, Sand~1, 72076 T\"ubingen, Germany
\and Hobby-Eberly Telescope, University of Texas at Austin, Austin, TX 78712, USA}

\date{Received 22 July 2024 \ Accepted 31 July 2024}

\abstract
{We present an analysis of 17 H-rich central stars of planetary nebulae (PNe) observed in our spectroscopic survey of nuclei of faint Galactic PNe carried out at the 10-m Hobby-Eberly Telescope. Our sample includes ten O(H) stars, four DAO white dwarfs (WDs), two DA WDs, and one sdOB star. The spectra were analyzed by means of NLTE model atmospheres, allowing us to derive the effective temperatures, surface gravities, and He abundances of the central stars. Sixteen of them were analyzed for the first time, increasing the number of hot H-rich central stars with parameters obtained through NLTE atmospheric modeling by approximately 20\%.
We highlight a rare hot DA WD central star, Abell~24, which has a \Teff\ likely in excess of 100\,kK, as well as the unusually high gravity mass of $0.70 \pm 0.05$\,\Msol\ for the sdOB star Pa~3, which is significantly higher than the canonical extreme horizontal-branch star mass of $\approx\!0.48\,\mathrm{M}_{\odot}$.
By investigating Zwicky Transient Facility light curves, which were available for our 15 northern objects, we found none of them show a periodic photometric variability larger than a few hundredths of a magnitude. This could indicate that our sample mainly represents the hottest phase during the canonical evolution of a single star when transitioning from an asymptotic giant branch star into a WD.
We also examined the spectral energy distributions, detecting an infrared excess in six of the objects, which could be due to a late-type companion or to hot ($\approx\! 10^3$\,K) and\slash or cool ($\approx$100\,K) dust. We confirm previous findings that spectroscopic distances are generally higher than found through \Gaia\/ astrometry, a discrepancy that deserves to be investigated systematically.} 


\maketitle

\section{Introduction}
\label{sec:intro}

About two-thirds of all central stars of planetary nebulae (CSPNe) have hydrogen-rich atmospheres, according to the catalog of CSPNe published by \cite{Weidmann+2020}. However, although this catalog contains 290 CSPNe classified as H-rich, the number that have actually been analyzed by means of non-local thermodynamic equilibrium (NLTE) model atmospheres is less than a hundred. A large sample of 27 old, H-rich CSPNe was analyzed by \cite{Napiwotzki1999} based on optical spectra and metal-free NLTE models. In addition, far-ultraviolet (FUV) spectra of 10 and 15 H-rich CSPNe were analyzed by \cite{HeraldBianchi2011} and \cite{ZieglerPhD2012}, respectively, using sophisticated metal-line-blanketed expanding and static NLTE models. Atmospheric parameters for further H-rich CSPNe have been  presented in works related to individual objects or as part of analyses from surveys targeting CSPNe and hot white dwarfs (WDs) and pre-WDs \citep{Herrero+1990, Liebert+1995, McCarthy+1997, Mendez+1988, Mendez+1992, Rauch+1999, DeMarco+2015, Aller+2015, Reindl+2014a, Reindl+2017, Reindl+2021, Reindl+2023, Hillwig+2017, Werner+2019, Jeffery+2023, BondAbell57_2024}. Based on a literature search, we count 75 H-rich hot CSPNe that have been analyzed previously using NLTE atmospheric models.

The prima facie assumption is that H-rich CSPNe represent the hottest and shortest-lived phase during the  canonical evolution of a single star when transitioning from an asymptotic giant branch (AGB) star into an H-rich WD\null.  However, the metamorphosis of an AGB or red-giant-branch (RGB) star into a WD remains far from being fully understood. We still do not even know whether every star actually goes through a planetary-nebula (PN) phase or whether PNe are mainly the outcome of binary interactions \citep{Bond2000, Moe+2006, Moe+2012, DeMarco+2009, Jones2019, Boffin+Jones+2019}.\\

In fact, the H-rich CSPNe include numerous examples that do not conform to the classical (single star) picture of evolution. \cite{Weidmann+2020} state that close binary CSPNe are preferably found among H-rich CSPNe. These binaries include reflection-effect systems consisting of a hot (pre-)WD and a strongly irradiated late-type companion (e.g., \citealt{Bond1978, Shimanskii+2008}). A few of these systems even raise the question of whether the occurrence of PNe is restricted to post-AGB stars or whether PNe could also be created around post-RGB stars \citep{Hall+2013, Hillwig+2016, Hillwig+2017, Jones+2020, Jones+2022}. In addition, there are a few H-rich double-degenerate binary central stars known (e.g., \citealt{DeMarco+2015, SG+2015, Reindl+2020}) as well as one system consisting of an H-rich pre-WD and a possible A-type companion with an orbital period of $\approx$3300\,d \citep{Aller+2015, Jones+2017}.

Furthermore, the H-rich group of CSPNe includes two examples of central stars that allow us to witness stellar evolution in ``real time.'' 
These objects, which are thought to have undergone a late He-shell flash, are \object{FG\,Sge} and \object{V839\,Ara (SAO\,244567)}. 
The object \object{FG\,Sge} was observed to be H-rich in the 1960s, but it turned into an H-deficient and $s$-process enriched object a few decades later \citep{JefferySchoenberner2006}. The object \object{V839\,Ara}, on the other hand, still shows an H-rich composition and is currently returning to the AGB---where in a couple of hundred years it will likely transform into an H-deficient star as well \citep{Reindl+2014a, Reindl+2017}.  

Some apparent PNe around hot H-rich stars turn out to actually be Str\"omgren spheres in the interstellar medium rather than material ejected from the star. The ionizing sources of these low-excitation ``PN mimics'' are often found to be He-core burning extreme-horizontal branch (EHB) stars \citep{Frew+2010, Hillwig+2022} or post-EHB stars \citep{Ziegler+2012}.

\smallskip

This is the sixth work in a series of papers presenting results from a spectroscopic survey of nuclei of faint Galactic PNe. The survey was carried out with the second generation Low-Resolution Spectrograph (LRS2; \citealt{Chonis2016}) of the 10-m Hobby-Eberly Telescope (HET; \citealt{Ramsey1998,Hill2021}) located at McDonald Observatory in west Texas, USA\null. An overview of the survey as well as a description of the instrumentation and data reduction procedures, target selection, and some initial results were presented in our first paper \citep[][hereafter Paper~I]{Bond2023a}. Paper~II \citep{Bond2023b} discusses the central star of the PN mimic Fr~2-30, and Paper~III \citep{WernerBondIII2024} presents discoveries of three new extremely hot H-deficient planetary-nebula nuclei (PNNi). In Paper~IV \citep{BondPa27_2024}, we reported that the nucleus of Pa~27 is a rapidly rotating late-type star, and Paper~V \citep{BondAbell57_2024} focuses on the peculiar central star of the PN, Abell~57. In this sixth publication, we present analyses of 17 hot and H-rich CSPNe. Of these, 16 were analyzed for the first time, increasing the number of hot H-rich CSPNe for which NLTE atmospheric parameters are available by $\approx$20\%.

In \se{sec:targets} we introduce the targets, and in \se{sec:observations} we give an overview of the observations. The spectral classification and analysis are presented in \se{sect:class} and \se{sect:analysis}, respectively. We then derive the Kiel masses (\se{sect:kiel}), search for photometric variability (\se{sect:lc}), and perform fits to the spectral energy distributions (SEDs) of our stars (\se{sect:seds}). We discuss our results in \se{sec:discussion} and conclude in \se{sec:con}.


\begin{table*}
\centering
\caption{PN target list and \Gaia\/ DR3 data for central stars.}
\label{tab:targetlist}
\begin{tabular}{lccccccccc}
\hline\hline
\noalign{\smallskip}
Name 
& PN G
& RA (J2000)
& Dec (J2000)
& $l$ [deg]
& $b$ [deg]
& Parallax [mas]
& {$G$}
& $G_{\rm BP}-G_{\rm RP}$
& $R_{\mathrm{PN}}$ \\
\noalign{\smallskip}
\hline
\noalign{\smallskip}
Abell 6   & 136.1+04.9    &  02 58 41.864 &    +64 30 06.28   &  136.10 & +04.93   & $ 0.869   \pm  0.143$ &  18.40 &  $ 0.97 $ & $\phantom{0}94''$ \\ 
Abell 16  & 153.7+22.8    &  06 43 55.418 &    +61 47 24.66   &  153.77 & +22.83   & $ 0.762   \pm  0.173$ &  18.62 &  $-0.18 $ & $\phantom{0}74''$ \\
Abell 24  & 217.1+14.7    &  07 51 37.554 &    +03 00 21.16   &  217.17 & +14.75   & $ 1.391   \pm  0.097$ &  17.37 &  $-0.57 $ & $198''$ \\
Abell 28  & 158.8+37.1    &  08 41 35.555 &    +58 13 48.38   &  158.80 & +37.18   & $ 2.606   \pm  0.071$ &  16.50 &  $-0.44 $ & $165''$\\
Abell 62  & 047.1$-$04.2  &  19 33 17.889 &    +10 36 59.76   &  047.18 & $-$04.29 & $ 0.186   \pm  0.213$ &  18.57 &  $-0.02 $ & $\phantom{0}83''$\\  
Alv 1     & 079.8$-$10.2  &  21 15 06.658 &    +33 58 18.74   &  079.89 & $-$10.26 & $ 0.604   \pm  0.127$ &  18.28 &  $-0.37 $ & $135''$ \\
Fe 4      & 030.6$-$16.4  &  19 46 30.838 &    $-$09 21 20.00 &  030.66 & $-$16.45 & $ 0.158   \pm  0.051$ &  16.07 &  $-0.23 $ & $\phantom{0}15''$ \\  
Kn 2      & 043.3+10.4    &  18 32 40.005 &    +13 58 02.43   &  043.40 & +10.41   & $ 0.108   \pm  0.079$ &  17.30 &  $-0.11 $ & $\phantom{0}28''$ \\  
Kn 40     & 198.6$-$06.7  &  06 00 47.193 &    +09 28 39.90   &  198.64 & $-$06.74 & $ 0.193   \pm  0.111$ &  17.46 &  $-0.11 $ & $18.5''$ \\ 
Kn 45     & 066.5$-$14.8  &  20 53 03.941 &    +21 00 10.96   &  066.51 & $-$14.90 & $ 0.484   \pm  0.175$ &  18.46 &  $-0.41 $ & $72.5''$ \\ 
Pa 3      & 090.8+06.1    &  20 46 10.639 &    +52 57 05.39   &  090.82 & +06.12   & $ 0.986   \pm  0.028$ &  15.90 &  $ 0.16 $ & $\phantom{0}36''$ \\ 
Pa 12     & 012.4+17.8    &  17 09 38.625 &    $-$09 00 41.47 &  012.46 & +17.84   & $ 0.045   \pm  0.081$ &  17.10 &  $ 0.45 $ & $\phantom{0}8.5''$ \\ 
Pa 15     & 064.9$-$09.1  &  20 29 07.586 &    +23 11 09.65   &  064.97 & $-$09.13 & $ 0.268   \pm  0.072$ &  16.67 &  $-0.19 $ & $\phantom{0}6.0''$ \\ 
Pa 26     & 056.9$-$11.7  &  20 19 46.164 &    +15 14 07.27   &  057.00 & $-$11.71 & $-0.082   \pm  0.074$ &  16.91 &  $-0.30 $ & $\phantom{0}6.5''$ \\ 
Pa 41     & 098.3$-$04.9  &  22 10 13.647 &    +50 04 33.41   &  098.31 & $-$04.93 & $ 0.469   \pm  0.072$ &  17.36 &  $-0.09 $ & $\phantom{0}51''$ \\ 
Pa 157    & 035.7+19.2    &  17 47 08.507 &    +11 00 20.96   &  035.72 & +19.20   & $ 0.308   \pm  0.054$ &  16.30 &  $-0.27 $ & $\phantom{0}15''$ \\ 
Pre 8     & 134.3$-$43.2  &  01 26 36.058 &    +18 51 18.36   &  134.38 & $-$43.23 & $ 0.215   \pm  0.152$ &  18.10 &  $-0.34 $ & $\phantom{0}48''$ \\ 
\noalign{\smallskip}
\hline\hline
\end{tabular} 
\end{table*}

\section{Targets}
\label{sec:targets}

As explained in our previous papers, we assembled a lengthy target list of central stars, most of them belonging to faint PNe discovered in recent years by amateur astronomers, and this list was submitted to the HET observing queue \citep[see][]{Shetrone2007PASP}. Exposures were chosen for execution by the telescope schedulers from the list, essentially at random, depending on sky conditions and lack of observable higher-ranked targets. 

The first two columns of Table~\ref{tab:targetlist} list the names and PN\,G designations of the host PNe for the 17 central stars for which we obtained the observations discussed in this paper. The next columns give the celestial and Galactic coordinates, parallaxes, and magnitudes and colors of the central stars, all taken from \Gaia\/ Data Release~3\footnote{\url{https://vizier.cds.unistra.fr/viz-bin/VizieR-3?-source=I/355/gaiadr3}} (DR3; \citealt{Gaia2016, Gaia2023}). Further information about the objects, including direct images of the PNe at several wavelengths, is contained in the online Hong-Kong/AAO/Strasbourg/H$\alpha$ Planetary Nebulae (HASH) database\footnote{\url{http://hashpn.space/}} \citep{Parker2016, Bojicic2017}. 
The final column in Table~\ref{tab:targetlist} gives the angular radii of the PNe, taken from HASH\null. Most of the nebulae are relatively large on the sky, but three are more compact, with radii of less than $10''$.

The following subsections give brief details of the discoveries of these faint PNe and their nuclei and some of the nebular properties. 
As noted in the introduction, the 17 stars analyzed in the present paper all have H-rich atmospheres. Separate papers in our series have analyzed H-deficient nuclei and other objects of special interest. To our knowledge, all but one of the 17 stars in Table~\ref{tab:targetlist} have not previously had their spectra discussed in the literature; they are not contained in the recent compilation of spectral classifications of central stars assembled by \citet{Weidmann+2020}. The one exception is the nucleus of Abell~28, discussed in detail below (Section~\ref{sect:analysis}).

\subsection{Abell 6, 16, 24, 28, and 62}

These five nebulae, Abell 6, 16, 24, 28, and 62, were discovered in the classical search of the Palomar Observatory Sky Survey (POSS) photographs for low-surface-brightness PNe by \citet{Abell1966}. Abell identified the faint blue central stars for four of them and presented photoelectric or photographic photometry. He was unable to locate the nucleus of Abell 62, likely because of the nebula's large angular size (diameter $166''$) and its superposition on an extremely rich star field. However, our inspection of the Space Telescope Science Institute's Digitized Sky Survey\footnote{\url{https://archive.stsci.edu/cgi-bin/dss_form}} (DSS) images from 2019 revealed a faint blue star near its center, which we added to our target list. All five Abell central stars are contained in a catalog of PNNi assembled from \Gaia\/ DR2 data by \citet{Chornay2020}. One or more of them are included in catalogs of PNNi constructed from searches of \Gaia\/ EDR3 for PNNi by \citet{Gonzalez2021} and/or for WDs by \citet{GentileFusillo2021}. Several are also contained in a catalog of central stars created by \citet{GomezMunoz2023}, who correlated PNe in the HASH database with UV images obtained by the Galaxy Evolution Explorer (GALEX) and other sky surveys.

\subsection{Alv 1}

This large (diameter $270''$) and very low-surface-brightness PN was discovered serendipitously by amateur Felipe Alves in deep narrow-band images he obtained of the nearby and much brighter nebula MWP~1. Details of the discovery of Alv~1 are given by \citet{Acker2012}, who presented images and pointed out its 18th-magnitude central star.\footnote{A deep amateur image of Alv~1 is posted at \url{http://www.capella-observatory.com/ImageHTMLs/PNs/MWP1andALV1.htm}} The nucleus is contained in several recent lists of hot subluminous stars and/or PNNi revealed in searches of the \Gaia\/ catalog, including those by \citet{Geier2019}, \citet{Chornay2020}, \citet{GentileFusillo2021},  and \citet{Culpan2022}.

\subsection{Fe 4}

The relatively bright central star of Fe~4 came to our attention through its inclusion in the catalog of PNNi prepared by \citet{Chornay2020}. The PN was discovered by Laurent Ferrero \citep{Ferrero2015}; it is a faint ring with a diameter of about $30''$, encircling a 16th-mag nucleus.\footnote{A deep narrow-band image of Fe~4 is available at \url{https://www.chart32.de/component/k2/objects/planetaty-nebulae/fe4-planetary-nebula}} The central star is cataloged as a hot subdwarf in the \Gaia-based lists of  \citet{Geier2019} and \citet{Culpan2022}. It is a conspicuous GALEX\/ source. 

\subsection{Kn 2, 40, and 45}

Kn 2, 40, and 45 were discovered by amateur Matthias Kronberger in his search of DSS images for faint PNe as part of the Deep Sky Hunters (DSH) collaboration \citep{Kronberger2006}. His discoveries of these three PNe were published by \citet{Jacoby2010}, who presented deep narrow-band images of the objects. All three central stars are listed in the \Gaia-based catalogs of hot subdwarfs assembled by \citet{Geier2019} and \citet{Culpan2022}, and Kn~40 and Kn~45 are also in the catalog of \citet{GentileFusillo2019}. However, only Kn~40 is contained in the \citet{Chornay2020} catalog of PNNi. The nuclei of Kn~2 and Kn~45 are bright GALEX\/ sources.

\subsection{Pa 3, 12, 15, 26, 41, and 157}
\label{subsect:pa3}

These six faint PNe (or candidate PNe) were discovered by Dana Patchick. Pa~3 was listed as a candidate PN by \citet{Kronberger2006}, and there is a narrow-band image of it in \citet{Jacoby2010}. The discovery of Pa~12, likewise accompanied by a narrow-band image, was published by \citet{Jacoby2010}. Pa~15, 26, and 41 were listed as new PNe by \citet{Kronberger2014}, and Pa~157 was included in a private communication of DSH discoveries to us from G.~Jacoby. 

All of the central stars are bright GALEX\/ sources, with the exception of Pa~12, which lies at a location never imaged by GALEX\null. The nuclei of Pa~3, 15, 26, and 157 are contained in the \citet{Geier2019} catalog of subluminous hot stars found in their \Gaia\/ survey, and those of Pa~3, 26, and 157 are listed by \citet{Culpan2022}.

\subsection{Pre 8}

Pre~8 is a high-Galactic-latitude PN discovered by Trygve Prestgard.\footnote{Details of Prestgard's discovery, and a deep narrow-band image, are available at \url{https://skyhuntblog.wordpress.com/planetary-nebulae/my-planetary-nebula-candidates/}} It is included in an extensive list of faint PNe discovered by French amateurs published by \citet{LeDu2018}. Our inspection of DSS images revealed its conspicuous blue central star. The nucleus is also listed as a hot subdwarf by \citet{Geier2019} and \citet{Culpan2022}.

\medbreak

\section{Observations and data reduction}

\label{sec:observations}

Paper~I gives full details of the LRS2 instrumentation used for our survey. We note here that LRS2 is composed of two integral-field-unit (IFU) spectrographs: blue (LRS2-B) and red (LRS2-R). All of the observations discussed in this paper were made with LRS2-B, which employs a dichroic beamsplitter to send light simultaneously into two units: the ``UV'' channel (covering 3640--4645~\AA\ at resolving power 1910) and the ``Orange'' channel (covering 
4635--6950~\AA\ at resolving power 1140). The data were initially processed using 
\texttt{Panacea},\footnote{\url{https://github.com/grzeimann/Panacea}} which performs bias and flat-field correction, fiber extraction, and wavelength calibration. An absolute-flux calibration comes from default response curves and measures of the telescope mirror illumination as well as the exposure throughput from guider images. We then applied \texttt{LRS2Multi}\footnote{\url{https://github.com/grzeimann/LRS2Multi}} to the un-sky-subtracted, flux-calibrated
fiber spectra in order to perform background and sky subtraction in an annular aperture, source extraction using a 2$\arcsec$ radius aperture, and combination of multiple exposures, if applicable, similar to the description in Paper~II\null. The final spectra from both channels were resampled to a common linear grid with a 0.7~\AA\ spacing and then normalized to a flat continuum for atmospheric analysis.  An observation log for our LRS2-B exposures is presented in 
Table~\ref{tab:observations}.

\begin{table} 
\caption{Log of HET LRS2-B observations.}
\label{tab:observations}
\begin{tabular}{lcl}
\hline 
\hline 
\noalign{\smallskip}
Central & Date & Exposure \\
Star of & [YYYY-MM-DD] & [s] \\
\hline
\noalign{\smallskip}
Abell 6   & 2019-12-01 & $3\times500   $ \\        
          & 2020-12-25 & $2\times1200  $ \\
          & 2021-11-07 & $2\times1200  $ \\
Abell 16  & 2020-11-18 & $2\times1000  $ \\
          & 2021-10-30 & $2\times1200  $ \\
Abell 24  & 2020-11-22 & $2\times500   $ \\
          & 2021-11-11 & $2\times600   $ \\
Abell 28  & 2019-10-22 & $180     $ \\
          & 2021-11-05 & $360     $ \\
          & 2022-04-17 & $360     $ \\
Abell 62  & 2021-07-08 & $2\times1000  $ \\
          & 2021-10-03 & $2\times1000  $ \\
          & 2022-07-20 & $2\times1200  $ \\
Alv 1     & 2022-06-10 & $2\times1000  $ \\
Fe 4      & 2023-07-19 & $240     $ \\
Kn 2      & 2019-10-06 & $2\times300   $ \\
Kn 40     & 2019-11-18 & $2\times300   $ \\
          & 2021-11-02 & $2\times600   $ \\
          & 2021-11-03 & $2\times600   $ \\
Kn 45     & 2019-10-30 & $60      $ \\
          & 2021-10-17 & $300     $ \\
Pa 3      & 2019-10-15 & $120     $ \\
          & 2022-06-11 & $360     $ \\
          & 2022-08-04 & $360     $ \\
Pa 12     & 2019-08-05 & $450     $ \\
Pa 15     & 2020-08-06 & $240     $ \\
          & 2021-05-20 & $500     $ \\
          & 2021-11-10 & $500     $ \\
          & 2022-05-08 & $2\times600   $ \\
          & 2022-05-15 & $2\times600   $ \\
Pa 26     & 2020-10-17 & $360     $ \\
Pa 41     & 2020-11-19 & $2\times375   $ \\
          & 2021-05-18 & $2\times750   $ \\
          & 2022-05-22 & $2\times750   $ \\
          & 2022-06-10 & $2\times750   $ \\
Pa 157    & 2021-02-24 & $200     $ \\
          & 2021-05-18 & $400     $ \\
          & 2022-03-21 & $400     $ \\
          & 2022-05-21 & $400     $ \\
Pre 8     & 2022-07-10 & $800     $ \\
\noalign{\smallskip}
\hline\hline
\end{tabular} 
\end{table}


\section{Spectral classification}
\label{sect:class}

Table~\ref{tab:paramaters} lists our 17 targets and their atmospheric parameters, helium contents, and masses derived in the discussion in the next section. The second column in the table lists spectral types inferred from our LRS2-B spectra. The majority (ten) of our targets are classified as O(H) stars, according to the classification scheme of \cite{Mendez1991}. A large fraction of central stars are assigned to this spectral type \citep{Weidmann+2020}. They are hot H-rich pre-WDs approaching their maximum effective temperature during post-AGB evolution. Six of our central stars are hot H-rich WDs. According to the usual classification scheme \citep[e.g.,][]{Wesemael1993}, four of them are DAO WDs because they display an ionized helium line (\Ionw{He}{2}{4686}). The other two are DA WDs, that is, they have a pure Balmer-line spectrum. Finally, we have one exceptional object (Pa~3) that displays a spectrum of a hot subdwarf. It is of spectral type sdOB, meaning that besides the Balmer lines, we observe lines from neutral helium plus \Ionw{He}{2}{4686} \citep[see, e.g., ][]{Heber2016}.

\begin{table*}
\centering
\caption{Spectral types, atmospheric parameters, and Kiel masses of our 17 H-rich CSPNe.}
\label{tab:paramaters}
\begin{tabular}{lcccccc}
\hline 
\hline 
\noalign{\smallskip}
Name & Spectral Type & \Teff\ [kK] & \logg & $\log \rm(He/H)$ & $M$ [$\mathrm{M}_{\odot}$] & Remarks\\
\hline
\noalign{\smallskip}
Abell 6   & DAO  & $ \phantom{0}86.9  \pm 6.8  $ & $ 6.90  \pm  0.12 $ & $ -1.35 \pm     0.13 $ & $0.50 \pm 0.04 $&  \\       
Abell 16  & DAO  & $ \phantom{0}83.4  \pm 1.5  $ & $ 7.27  \pm  0.06 $ & $ -1.60 \pm     0.06 $ & $0.54 \pm 0.02 $&  \\
Abell 24  & DA   & $ \phantom{0}109.2 \pm 1.0  $ & $ 7.70  \pm  0.03 $ & $ \dots              $ & $0.66 \pm 0.03 $& (1) \\
Abell 28  & DA   & $ \phantom{0}63.8  \pm 0.4   $ & $ 7.76  \pm  0.02 $ & $ \dots              $ & $0.61 \pm 0.02 $& (1) \\
Abell 62  & DAO  & $ \phantom{0}89.1  \pm 1.7  $ & $ 7.46  \pm  0.05 $ & $ -1.75 \pm     0.06 $ & $0.58 \pm 0.03 $&  \\
Alv 1     & O(H) & $ \approx170.0    $ & $ \approx6.55 $ & $\approx -0.71$ & $ \approx 0.62 $& (2) \\
Fe 4      & O(H) & $ \phantom{0}75.7  \pm 2.2  $ & $ 5.25  \pm  0.02 $ & $ -1.37 \pm     0.04 $ & $0.51 \pm 0.02 $&  \\
Kn 2      & O(H) & $ \phantom{0}85.7  \pm 2.4  $ & $ 5.33  \pm  0.04 $ & $ -1.03 \pm     0.06 $ & $0.55 \pm 0.04 $&  \\
Kn 40     & O(H) & $ \phantom{0}95.2  \pm 2.1  $ & $ 5.71  \pm  0.02 $ & $ -1.36 \pm     0.04 $ & $0.53 \pm 0.05 $&  \\
Kn 45     & DAO  & $ \phantom{0}90.2  \pm 2.3  $ & $ 6.86  \pm  0.06 $ & $ -1.55 \pm     0.07 $ & $0.50 \pm 0.02 $&  \\
Pa 3      & sdOB & $ \phantom{0}37.2  \pm 0.1   $ & $ 6.11  \pm  0.01 $ & $ -1.56 \pm     0.02 $ & $\approx 0.48$& (3) \\
Pa 12     & O(H) & $ \phantom{0}65.0  \pm 5.8  $ & $ 5.12  \pm  0.07 $ & $ -1.44 \pm     0.08 $ & $0.45 \pm 0.04 $& (4) \\
Pa 15     & O(H) & $ \approx100.0  $ & $ \approx5.18 $ & $ \approx-1.05 $ & $\approx 0.67 $& (2,4) \\
Pa 26     & O(H) & $ \phantom{0}94.1  \pm 3.6  $ & $ 5.16  \pm  0.02 $ & $ -1.01 \pm     0.04 $ & $0.63 \pm 0.04 $& (4) \\
Pa 41     & O(H) & $ 136.6 \pm 6.6  $ & $ 5.95  \pm  0.02 $ & $ -0.78 \pm         0.04 $ & $0.61 \pm 0.04 $&  \\
Pa 157    & O(H) & $ \phantom{0}91.0  \pm 0.9   $ & $ 5.44  \pm  0.01 $ & $ -1.07 \pm     0.02 $ & $0.55 \pm 0.02 $&  \\
Pre 8     & O(H) & $ 130.0 \pm 9.6  $ & $ 6.44  \pm  0.07 $ & $ -0.80 \pm         0.10 $ & $0.55 \pm 0.05 $&  \\
\hline
\end{tabular} 
\tablefoot{  
\tablefoottext{1}{Pure H atmosphere;} 
\tablefoottext{2}{Parameters uncertain due to star lying near border of theoretical grid;}  
\tablefoottext{3}{{\tt sdOstar2020} grid used for analysis (see text);} 
\tablefoottext{4}{H$\alpha$ not included in fit due to oversubtraction of PN emission (see text). The helium abundances are given as logarithmic number ratios relative to H.} 
}
\end{table*}

\begin{figure*}
\centering
\includegraphics[width=\textwidth]{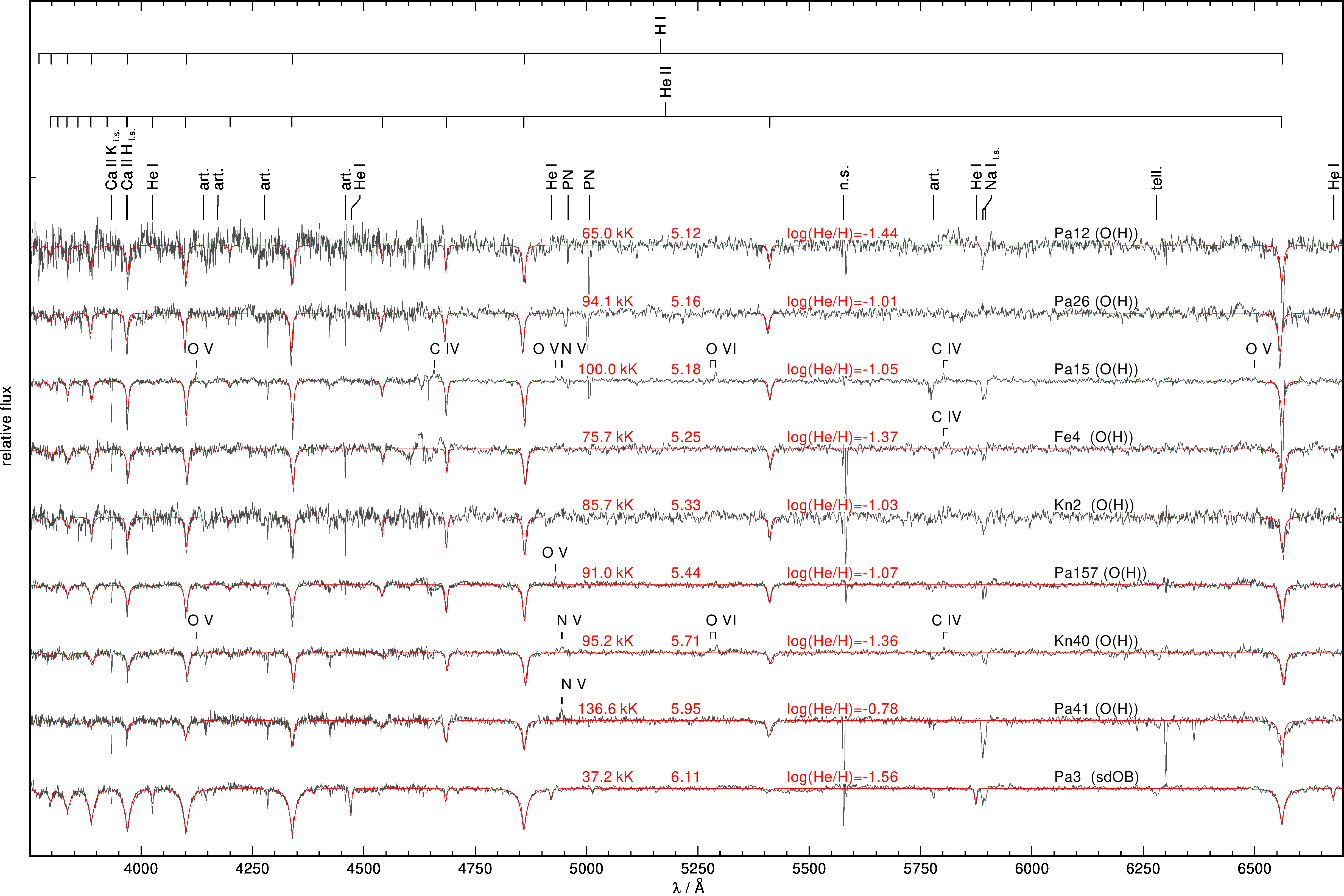}
\caption{
Model-atmosphere fits (red graphs) to observed spectra (black graphs) of nine of our H-rich central stars. The spectra are sorted by increasing $\log g$ from top to bottom. Spectral classifications, names of stars, derived effective temperatures, surface gravities, and logarithmic He/H ratios (by number) are indicated. Identified lines are marked. The PN emission lines are oversubtracted in the top three spectra. Features near 4640\,\AA\ are artifacts where the UV and Orange spectral channels are joined.  Several apparently sharp features are instrumental artifacts, marked ``art.'' Interstellar absorption lines are labeled ``i.s.," telluric absorption is marked ``tell.,'' and imperfectly subtracted night-sky features are labeled ``n.s.'' Oversubtracted nebular lines of [\ion{O}{iii}] in the top three spectra are marked ``PN.''
\label{fig:spectra_a}
}
\end{figure*}

\begin{figure*}
\centering
\includegraphics[width=\textwidth]{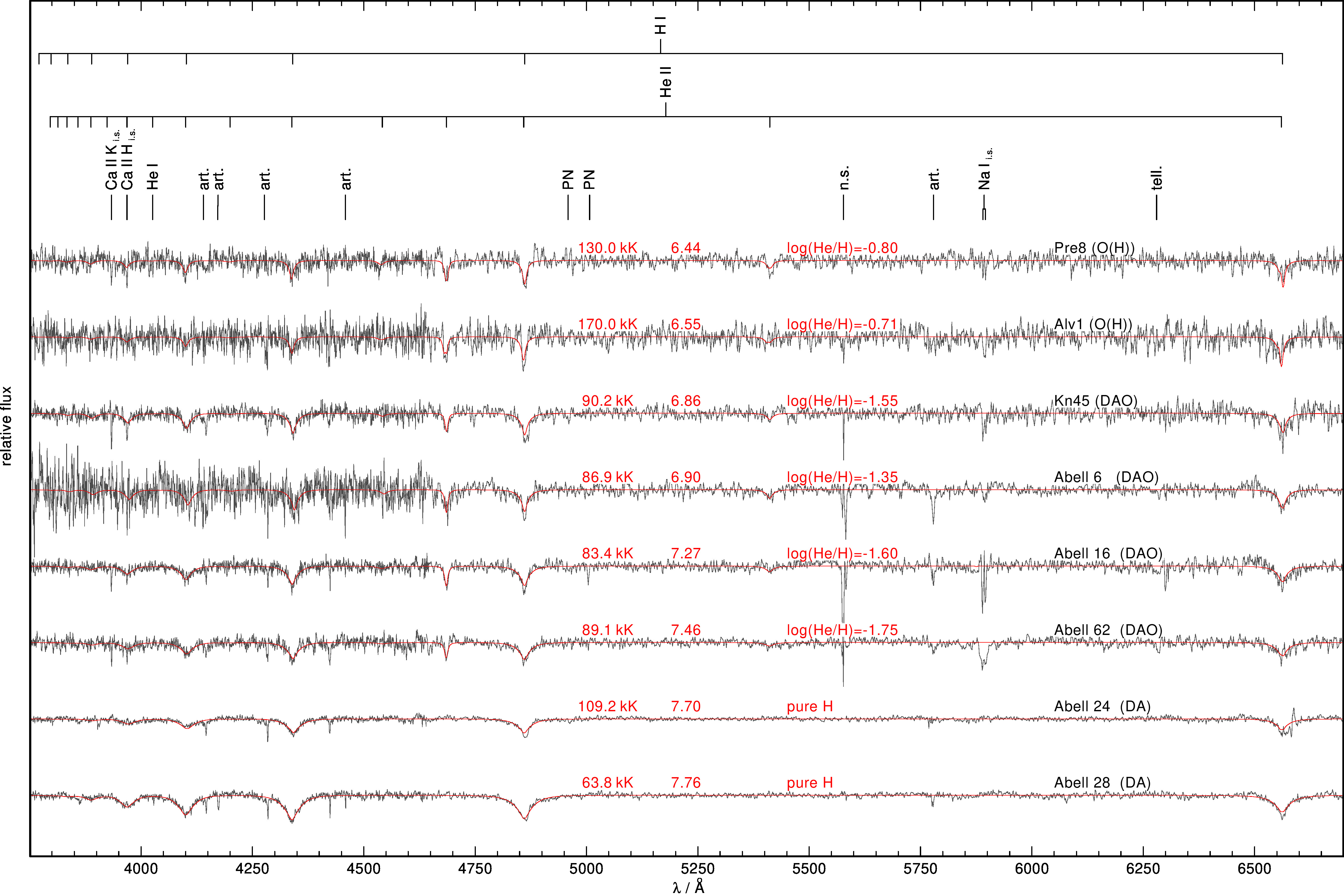}
\caption{
Similar to Figure~\ref{fig:spectra_a} but for the other eight central stars. 
\label{fig:spectra_b}
}
\end{figure*}

Figures~\ref{fig:spectra_a} and~\ref{fig:spectra_b} present plots of our rectified LRS2-B spectra. As shown in Figure~\ref{fig:spectra_a}, a few of our O(H) stars display very weak photospheric emission lines of highly ionized carbon, nitrogen, and\slash or oxygen. We identified \Ionw{C}{4}{4658} (in Pa~15), \Ionww{C}{4}{5801, 5812} (Fe~4, Kn~40, Pa~15), \Ionw{N}{5}{4945} (Kn~40, Pa~15, Pa~41), \Ionw{O}{5}{4124} (Kn~40, Pa~15), \Ionw{O}{5}{4930} (Kn~40, Pa~15, Pa~157), \Ionw{O}{5}{6500} (Pa~15), as well as \Ionw{O}{6}{5291} (Kn~40, Pa~15).

We note that the subtraction of emission lines from the surrounding PNe has succeeded well in the majority of cases, because most of the nebulae are extremely faint. Only for Pa~12, Pa~15, and Pa~26, whose spectra are the top three plotted in Figure~\ref{fig:spectra_a}, did we observe that the [\ion{O}{iii}] and H$\alpha$ emission lines have been oversubtracted. These targets are those with the three smallest PN angular radii (see Table~\ref{tab:targetlist}).

The unique assignment of central-star spectra to a spectral class often requires knowledge of the atmospheric parameters. In particular, O(H) stars and hot subdwarfs can have rather similar spectra. Also, the transition between the hottest O(H) stars and DAO WDs is continuous. One possibility to distinguish between both classes is the surface gravity. Somewhat arbitrarily, the value $\logg = 7$ is usually chosen, and above this value, a star is regarded as a WD \citep[e.g.,][]{Wesemael1993}. Perhaps more meaningful is the choice that we followed to call a star an O(H) type before it reaches the maximum effective temperature during its post-AGB evolution and to regard it as a (DAO) WD afterward, when it starts to cool along the WD sequence \citep[e.g.,][]{Reindl+2023}. In practice this means that the value of the surface gravity that limits O(H) stars from WDs is smaller than $\logg = 7$ and is around $\logg=6.4$ for the objects with the lowest post-AGB remnant masses.

\section{Atmospheric analysis}
\label{sect:analysis}

To derive the effective temperatures, surface gravities, and helium abundances for our program stars, we performed global $\chi^2$ spectral fits to consider several absorption lines of H and He and calculated the statistical 1$\sigma$ errors.  
For the DA- and DAO-type WDs, we used the metal-free model grids introduced in \cite{Reindl+2023}, while for the O(H)-type stars, we employed the H+He model grid of \cite{Reindl+2016}. For the sdOB star, Pa~3, we relied on the {\tt sdOstar2020} grid \citep{Dorsch+2021}, which also considers metal opacities for C, N, O, Ne, Mg, Al, Si, P, S, Ar, Ca, Fe, and Ni. All model atmospheres are hydrostatic, chemically homogeneous, and assume NLTE\null.
The derived atmospheric parameters are listed in columns 3 and 4 in Table~\ref{tab:paramaters}, and the helium-to-hydrogen ratios by numbers are in column 5. Our best spectral fits are plotted in Figures~\ref{fig:spectra_a} and~\ref{fig:spectra_b}.

For two stars (Pa~15 and Alv~1), the fit reached the border of the model grid; therefore, the derived parameters must be regarded as estimates only. As can be seen in Figures~\ref{fig:spectra_a} and~\ref{fig:spectra_b}, both stars suffer from a well-known Balmer-line problem, namely, the failure to achieve a consistent fit to all Balmer (and \Ion{He}{2}) lines simultaneously. This means that for a particular object, different values of $\Teff$ follow from fits to different Balmer line series members. This problem is commonly seen in very hot \mbox{(pre-)} WDs (e.g., \citealt{ Werner+2019}). We emphasize that generally, the Balmer-line analysis of the hottest H-rich \mbox{(pre-)}WDs is notoriously prone to large errors. A way out of this problem is the analysis of metal lines in  UV spectra. For example, the temperature of one of the hottest known DAs (PG\,0948$+$534) was estimated to be 140$\pm$12\,kK from a Balmer-line fit \citep{Preval2017}, but a UV analysis revealed a significantly lower temperature of 105$\pm$5\,kK \citep{Werner+2019}. In addition, we speculate that since Pa\,15 is close to the Eddington limit, it likely still has a weak stellar wind. Thus, models for expanding atmospheres would be more suitable for the analysis of this star. 

For Pa~12, Pa~15, and Pa~26, we excluded H$\alpha$ from the fit, as it suffered from an oversubtraction of the nebular emission line. We note that correctly subtracting the nebular emissions can be challenging in the case of very compact and/or asymmetrical nebulae; therefore, systematic errors on the derived parameters need to be taken into account. For Pa~26---as an example---we derived $\Teff=92.5\pm3.5$\,kK and $\logg = 5.15\pm0.03$ when including H$\alpha$ in the fit and $\Teff = 94.1\pm3.6$\,kK and $\logg=5.16\pm0.02$ when excluding H$\alpha$. 

For the DA-type WD Abell~28, we derived $\Teff = 63.8\pm0.4$\,kK and $\logg = 7.76\pm0.02$ using pure-H models. This star also has two spectra that were obtained by the Sloan Digital Sky Survey (SDSS), and they have been analyzed in several previous studies, as listed in Table~\ref{tab:abell28}. Fitting the SDSS spectrum with the higher signal-to-noise ratio and using pure-H NLTE models, \cite{Kepler2019}, \cite{Tremblay2019}, and \cite{Bedard2020} derived $\Teff\approx57-58$\,kK and $\log g\approx7.7$.\footnote{We note that the lower S/N spectrum was fitted by \citet{Kleinman2013}, who derived $\Teff=62.4\pm0.1$\,kK and $\logg=7.58\pm0.05$. However, in this work pure-H LTE models were used; therefore, the slightly higher $\Teff$ compared to those reported by \cite{Kepler2019}, \cite{Tremblay2019}, and \cite{Bedard2020} might be a NLTE effect.} The SDSS spectra were obtained with a fiber diameter of 3$\arcsec$ and are very sensitive to faint nebular emission lines, even in the case of an evolved or a nearby nebula \citep{Yuan+2013}. Therefore, the detection of weak nebular lines in the SDSS spectra of Abell~28 (in particular in H$\alpha$, but also lines [\Ion{O}{3}] and [\Ion{N}{2}], are visible) might not be a surprise. 
We conclude that our derived atmospheric parameters are more reliable and that our higher $\Teff$ compared to the mentioned works is a result of poor background subtraction of the SDSS spectra.

\begin{table}
\centering
\caption{Stellar parameters for the nucleus of Abell~28. \label{tab:abell28} }
\begin{tabular}{cccc}
\hline 
\hline 
\noalign{\smallskip}
$\Teff$\,/\,kK & $\logg$ & Spectrum & Reference \\
 &  & Source &  \\
\hline 
\noalign{\smallskip}
$62.4\pm1.1$  & $7.58\pm0.05$  & SDSS$^{(a)}$ & $^{c)}$\\
$57.8\pm0.4 $ & $7.67\pm0.03$ & SDSS$^{(b)}$ & $^{(d)}$   \\
$57.2\pm0.7 $ & $7.66\pm0.05$ & SDSS$^{(b)}$ & $^{(e)}$  \\
$56.7\pm0.7 $ & $7.65\pm0.04$ & SDSS$^{(b)}$ & $^{(f)}$   \\
$63.8\pm0.4 $ & $7.76\pm0.02$ & HET    & $^{(g)}$        \\
\hline
\end{tabular} 
\tablefoot{  
\tablefoottext{a}{MJD 54425 spectrum} 
\tablefoottext{b}{MJD 56220 spectrum}  
\tablefoottext{c}{\cite{Kleinman2013} } 
\tablefoottext{d}{\cite{Kepler2019}}  
\tablefoottext{e}{\cite{Tremblay2019}} 
\tablefoottext{f}{This paper}  
\tablefoottext{g}{This paper}  
}
\end{table}

\begin{figure*}
\sidecaption
\includegraphics[width=12cm]{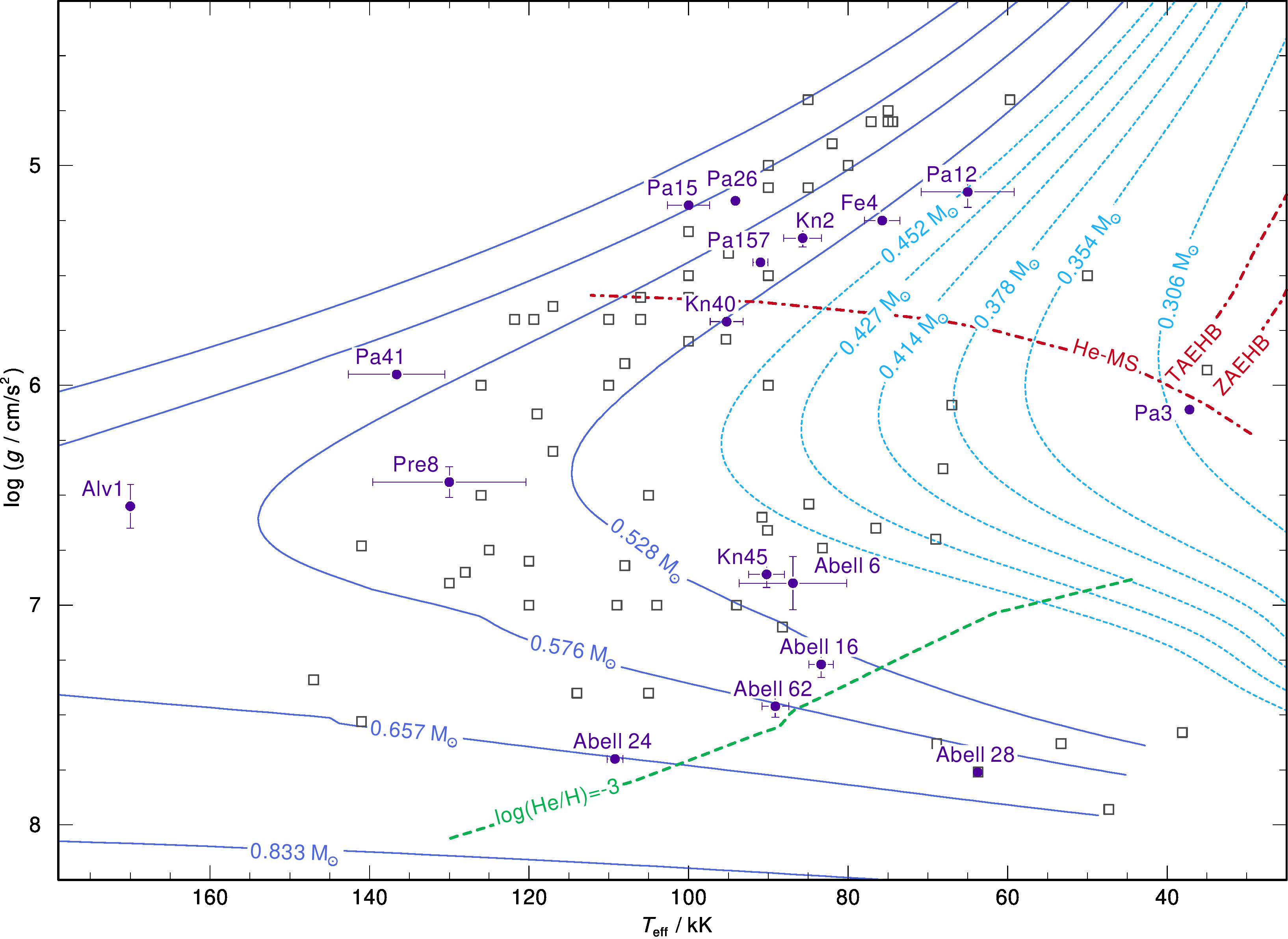}
\caption{
{Kiel diagram showing the positions of our H-rich central stars as purple filled circles. The gray open squares indicate previously analyzed H-rich CSPNe \citep{Herrero+1990, Liebert+1995, McCarthy+1997, Mendez+1988, Mendez+1992, Rauch+1999, DeMarco+2015, Aller+2015, Reindl+2014a, Reindl+2017, Reindl+2021, Reindl+2023, Hillwig+2017, Werner+2019, Jeffery+2023, BondAbell57_2024}. The blue solid lines are post-AGB evolutionary tracks from \cite{M3B2016}, and the light-blue dashed lines are post-RGB evolutionary tracks from \cite{Hall+2013}. The red dashed-dotted lines correspond to the He-burning main-sequence as well as the zero-age and terminal-age extended horizontal branches. The green dashed line indicates where the He abundances should have decreased down to $\log\mathrm{(He/H)} = -3$ according to predictions of \cite{UnglaubBues2000}.}
\label{fig:evolution}
}
\end{figure*}

\section{Kiel masses}
\label{sect:kiel}

In Fig.~\ref{fig:evolution} we show a Kiel diagram ($\log g$ versus $\Teff$) for our program stars along with post-AGB evolutionary tracks with metallicity $Z=0.02$ from \cite{M3B2016} and post-RGB tracks from \cite[][ priv.comm.]{Hall+2013}. 
For each star in our sample, we derived the stellar mass, $M$, using the theoretical evolutionary sequences and the {\tt griddata} interpolation function in \textsc{python}. Uncertainties of the masses were estimated using a Monte Carlo method. The derived 
Kiel masses are listed in column~6 of Table~\ref{tab:paramaters}.


\section{Light curves}
\label{sect:lc}

A significant fraction ($\approx$12--21\%) of Galactic CSPNe are photometrically variable \citep[e.g.,][]{Bond2000, Miszalski2009, Jacoby+2021}. This variability is often interpreted as originating from a close binary companion (e.g., \citealt{Boffin+Jones+2019}), but the variability can also stem from pulsations of the CS or in wide binary systems, from a cool spotted or pulsating companion \citep{Jacoby+2021, BondPa27_2024}. In addition, it has recently been proposed that a meaningful number of the hottest WDs develop spots on their surfaces when entering the WD cooling sequence \citep{Reindl+2021, Reindl+2023}.

In order to check if our stars show photometric variability, we inspected light curves from the Zwicky Transient Facility (ZTF; \citealt{Bellm+2019, Masci+2019}) survey data release 21, which provides photometry in the $g$ and $r$ bands and---less often---in the $i$ band. 
All of our 15 northern targets have publicly available light curves with around 50--1000 data points in the $g$ and $r$ bands. In Table~\ref{tab:ztf}, we provide an overview of the number of data points, mean magnitudes, and mean uncertainties for each object in our sample with available ZTF light curves.

For the analyses of the light curves, we used the \textsc{VARTOOLS} program 
\citep{HartmanBakos2016} to perform a generalized Lomb-Scargle (LS) search
\citep{Press1992, ZechmeisterKuerster2009} for periodic sinusoidal signals.
We consider objects variable if they show a periodic signal with a false-alarm 
probability (FAP) of $\log(FAP)\leq-4$.
Besides suspicious periods very close to one day, we did not find any hint of periodic variability in our stars. The typical uncertainty of the ZTF $g$-band light curves ranges from $0.01-0.02$\,mag for stars brighter than 18\,mag and up to 0.06 mag for the faintest stars in our sample; therefore a photometric variability larger than a few hundredths of a magnitude can be excluded.

The standard deviations of the g- and r-band measurements typically agree with the mean uncertainties. Only in the cases of Kn~45 and Pa~15 do the standard deviations in the $r$($g$)-band measurements significantly exceed the mean uncertainties by factors of 7.1(4.6) and 5.9(1.9) for both stars, respectively. We believe this is likely due to nearby and bright stars. Kn~45 has a nearby $G=11$\,mag star (\object{Gaia DR3 1814597642176881280}) at a distance of $11''$, and Pa~15 has a $G=13$\,mag star (\object{Gaia EDR3 1830887108803635584}) at a distance of $4''$. We also note that we did not detect any hints of long-term trends in the light curves, which cover approximately six~years.

\begin{table}
\centering
\caption{Names, bands of the light curves, number of data points, mean magnitudes, and mean uncertainties of the objects in our sample with available ZTF light curves. \label{tab:ztf} }
\begin{tabular}{lcccc}
\hline 
\hline 
\noalign{\smallskip}
Name    &       Band    &       Data    &       Magnitude &     Mean error      \\
        &               &       points  &       [mag]   &       [mag]   \\
\hline 
\noalign{\smallskip}
Abell~6 &       g       &       492     &       18.94   &       0.05    \\
        &       r       &       691     &       18.38   &       0.03    \\
Abell~16        &       g       &       333     &       18.51   &       0.05    \\
        &       r       &       436     &       18.87   &       0.06    \\
Abell~24        &       g       &       221     &       17.14   &       0.02    \\
        &       r       &       480     &       17.72   &       0.02    \\
        &       i       &       34      &       18.15   &       0.04    \\
Abell~28        &       g       &       527     &       16.34   &       0.02    \\
        &       r       &       785     &       16.79   &       0.02    \\
        &       i       &       68      &       17.18   &       0.03    \\
Abell~62        &       g       &       502     &       18.62   &       0.06    \\
        &       r       &       892     &       18.91   &       0.06    \\
        &       i       &       75      &       19.06   &       0.08    \\
Alv~1   &       g       &       486     &       18.14   &       0.03    \\
        &       r       &       973     &       18.59   &       0.04    \\
Kn~2    &       g       &       379     &       17.25   &       0.02    \\
        &       r       &       944     &       17.51   &       0.02    \\
Kn~40   &       g       &       352     &       17.40   &       0.02    \\
        &       r       &       532     &       17.68   &       0.02    \\
        &       i       &       36      &       18.00   &       0.03    \\
Kn~45   &       g       &       461     &       18.21   &       0.03    \\
        &       r       &       909     &       18.34   &       0.04    \\
        &       i       &       101     &       18.36   &       0.06    \\
Pa~3    &       g       &       58      &       15.99   &       0.01    \\
        &       r       &       83      &       16.03   &       0.01    \\
Pa~15   &       g       &       52      &       16.36   &       0.02    \\
        &       r       &       98      &       16.76   &       0.02    \\
Pa~26   &       g       &       80      &       16.72   &       0.02    \\
        &       r       &       124     &       17.11   &       0.02    \\
Pa~41   &       g       &       573     &       17.23   &       0.02    \\
        &       r       &       806     &       17.44   &       0.02    \\
        &       i       &       47      &       17.57   &       0.02    \\
Pa~157  &       g       &       93      &       16.20   &       0.01    \\
        &       r       &       140     &       16.56   &       0.01    \\
Pre~8   &       g       &       125     &       17.87   &       0.03    \\
        &       r       &       152     &       18.29   &       0.03    \\
        &       i       &       82      &       18.50   &       0.05    \\
\hline
\end{tabular} 
\end{table}

\section{Spectral energy distributions}
\label{sect:seds}

We performed fits to the SEDs of each star, employing the
$\chi ^2$ SED fitting routine described in \cite{Heber+2018} and 
\cite{Irrgang+2021}. For this purpose, we collected photometry 
from various catalogs, such as GALEX\/ \citep{Bianchi+2017}, Pan-STARRS \citep{Flewelling+2020}, 
Johnson \citep{Henden+2015}, \Gaia\/ (cyan, \citealt{Gaia+2020, Gaia+2021}), 2MASS \citep{Cutri2003}, SDSS \citep{Alam+2015}, Skymapper \citep{Onken+2019}, and in the fit, 
we kept the atmospheric parameters fixed and let the angular diameter, $\Theta$, and the color excess, 
$E(44-55)$,\footnote{\cite{Fitzpatrick+2019} employed $E(44-55)$, which is the monochromatic 
equivalent of the usual $E(B-V)$, using the wavelengths 4400\,\AA\ and 5500\,\AA, respectively. 
For high effective temperatures, such as for the stars in our sample, $E(44-55)$ is identical to $E(B-V)$.} vary freely. Interstellar reddening was accounted for by using the reddening law of \cite{Fitzpatrick+2019} with $R_V=3.1$.

\begin{figure*}
\centering
\includegraphics[width=\textwidth]{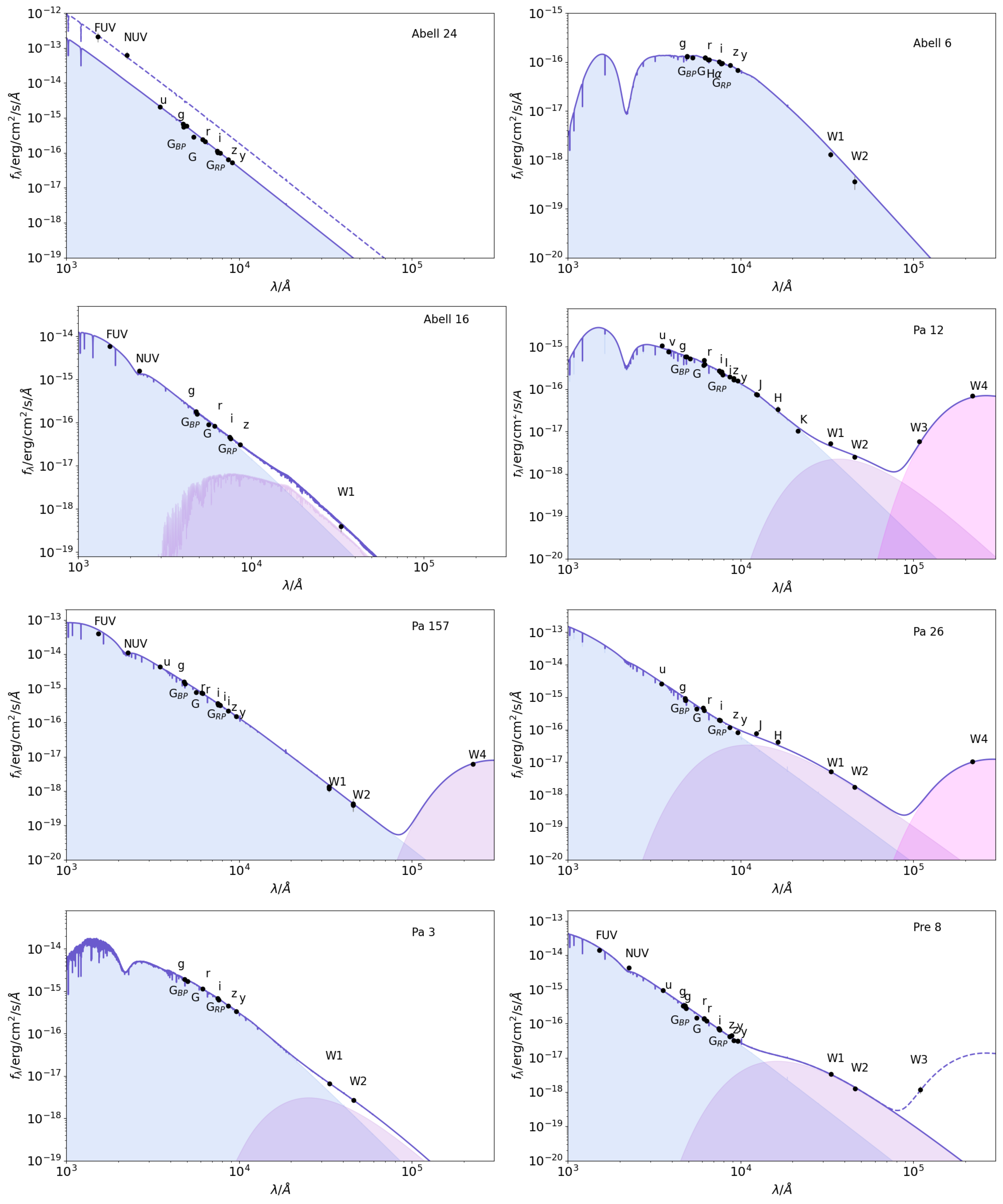}
\caption{
Spectral energy distribution fits for eight central stars. Top row: Two examples of SED fits for stars that do not show an IR excess. Remaining spectra: Multicomponent SED fits for the  stars in our sample that show a near- and\slash or far-infrared excess. Black dots correspond to observed photometry; the blue and purple shaded areas correspond to the flux contribution of the CS and a cool blackbody, respectively; and the dark purple line is the (combined) best fit to the observation.
\label{fig:seds}
}
\end{figure*}

In the top row of Fig.~\ref{fig:seds}, we show two examples of SED fits for stars that do not show an infrared excess.
For Abell\,24, we noticed that the GALEX\/ FUV and NUV fluxes exceed the flux that is 
predicted by optical photometry by a factor of five, which is indicated by the dark purple dashed line in Fig.~\ref{fig:seds}.
The origin of this UV excess is not understood, and we consequently excluded the GALEX\ fluxes from the SED
fit. The other example is the SED for Abell\,6, for which we found the highest color excess:
$E(44-55)= 0.983$\,mag.

We found that four of the O(H) stars, one DAO WD, and the sdOB star show a near- and\slash or far-infrared excess.
The SEDs of those objects are also displayed in Fig.~\ref{fig:seds}.
For those objects, we carried out a multicomponent fit using our best-fit model flux from the atmospheric 
analysis and one or two blackbody components. In those fits, we additionally let the blackbody $\Teff$ and 
surface ratios of the blackbodies relatively to that of the star vary freely. We note that for the 
{Wide-Field Infrared Survey Explorer} ({\it WISE}; \citealt{Wright+2010}) bands W1 and W2, we relied on 
magnitudes reported in the unWISE catalog \citep{Schlafly+2019}, as it is based on significantly deeper imaging and 
offers improved modeling of crowded regions compared to the ALLWISE catalog \citep{Cutri+2014}.

Pa\,3 shows an excess in the {\it WISE\/} W1 and W2 bands, which can be reproduced with a blackbody with $\Teff=1180$\,K and a surface ratio of 140 relative to that of the sdOB star.

For Pa\,157 the far-infrared excess shows up only in the W4 band, which can be modeled with a blackbody with $\Teff=95$\,K and a large surface ratio of $10^9$ relative to that of the O(H) star. For Pa\,12, two blackbody components---one with 800\,K and a surface ratio of $3100$ relative to that of the O(H) star---and another component---with 110\,K and a large surface ratio of $2\times10^9$---are needed to reproduce the far-infrared excess in all four {\it WISE\/} bands.
For Pa\,26, the infrared excess shows up already at around 10\,000\,\AA. The SED of Pa\,26 can be reproduced by assuming our best fit model for the O(H) star, two blackbodies with $\Teff=2690$\,K and $\Teff=100$\,K, and surface ratios relative to the O(H) star of 640 and $3\times10^9$, respectively. 

The O(H) star Pre\,8 shows a clear far-infrared excess in the {\it WISE\/} W1 and W2 bands, which can be modeled with a blackbody with $\Teff=1810$\,K and a surface ratio of 3700 relative to that of the O(H) star. There is also a possible excess in the W3 band, which would require another blackbody with $\Teff=110$\,K and a large surface ratio of $5.7\times10^{10}$, as indicated by the dark purple dashed line in Fig.~\ref{fig:seds}. We remain skeptical about the latter excess, since the S/N in the W3 band is only five, and---in contrast to the other stars with W3/W4 excesses---there is no detection in the W4 band.

For Abell~16, \cite{DeMarco+2013} have already reported their discovery of an $I$-band excess at the 2$\sigma$ level using high-precision photometry. They concluded that it is due to an M3\,V companion. Consequently, for this object we used PHOENIX models calculated by \cite{Husser+2013} and that cover effective temperatures between 2300\,K and 12000\,K instead of a blackbody component in order to reproduce the infrared excess. We find for the cool companion $\Teff=4200^{+1000}_{-600}$\,K and a surface ratio relative to that of the DAO star of $58^{+3}_{-25}$. 

We note that we do not expect a significant contribution of the nebulae to the optical and UV fluxes of the central stars. Otherwise, we would see excesses in the $g$, $r$, and/or \Gaia\/ bands, where the strongest nebular lines of [\Ion{O}{3}] and H$\alpha$ are located. This is not the case even for our most compact PNe.

For sources that have a \emph{Gaia} parallax measurement, 
$\varpi$, with a relative error on the 
parallax of 40\% or smaller, we calculated the radius, $R$, from the angular diameter via 
$R = \Theta/(2\varpi)$ and derived the star's luminosity from the radius and the 
$T_\mathrm{eff}$ from our spectral fitting via 
$L/L_\odot = (R/R_\odot)^2(T_\mathrm{eff}/T_{\mathrm{eff},\odot})^4$. 
Finally, we calculated the gravity mass from $M_{\mathrm{grav}} = g R^2/G$ from 
the radius and the $\logg$ as determined from the SED fitting and spectral analysis.
The derived radii, luminosities, and gravity masses\footnote{The numbers given are the median 
and the highest density interval with probability 0.6827 
\citep[see][for details on this measure of uncertainty]{Bailer-Jones+2021}.} are listed in Table~\ref{tab:SEDres}.

\begin{table*}
\centering
\caption{Values for the reddening for $E(B-V)$ from \cite{Schlafly2011} and $E(44-55)$ as determined in our SED fits, distances, 
zero-point corrected \emph{Gaia} distances, spectroscopic distances and radii, luminosities, and gravity masses. \label{tab:SEDres} }
\begin{tabular}{llllllll}
\hline 
\hline 
\noalign{\smallskip}
Name
& $E(B-V)$/mag
& $E(44-55)$/mag
& $d_{\mathrm{Gaia}}$/kpc
& $d_{\mathrm{spec}}$/kpc
& $R/R_\odot$ 
& $M_{\mathrm{grav}}/M_\odot$
& $L/L_\odot$ \\
\hline
\noalign{\smallskip}
Abell 6 & $1.090 \pm 0.060$     & $0.986 \pm 0.013$     & $1.07^{+0.21}_{-0.15}$        & $0.88^{+0.15}_{-0.13}$  & $0.053^{+0.011}_{-0.008}$     & $0.81^{+0.47}_{-0.29}$        & $140^{+90}_{-60}$       \\
\noalign{\smallskip}
Abell 16$^{1}$& $0.131 \pm 0.004$       & $0.157 \pm 0.015$     & $1.21^{+0.36}_{-0.23}$        & $2.02^{+0.17}_{-0.16}$  & $0.019^{+0.006}_{-0.004}$     & $0.24^{+0.18}_{-0.09}$        & $15^{+11}_{-\phantom{0}6}$      \\
\noalign{\smallskip}
Abell 24& $0.033 \pm 0.003$     & $0.007^{+0.016}_{-0.007}$     & $0.71^{+0.06}_{-0.05}$        & $0.96 \pm 0.05$ & $0.0143^{+0.0012}_{-0.0010}$  & $0.37^{+0.07}_{-0.06}$        & $26^{+5}_{-4}$  \\
\noalign{\smallskip}
Abell 28& $0.073 \pm 0.003$     & $0.067 \pm 0.005$     & $0.38\pm0.01$ & $0.420^{+0.013}_{-0.012}$       & $0.0156 \pm 0.0005$   & $0.51 \pm 0.04$       & $3.64^{+0.24}_{-0.22}$  \\
\noalign{\smallskip}
Abell 62& $0.506 \pm 0.009$     & $0.333 \pm 0.012$     &       & $1.28 \pm 0.09$   &       &       &       \\
\noalign{\smallskip}
Alv 1$^{2}$     & $0.127 \pm 0.004$     & $0.125 \pm 0.009$     & $1.58^{+0.44}_{-0.29}$        & $\approx6.50$   & $\approx0.018$        & $\approx0.05$ & $\approx250$  \\
\noalign{\smallskip}
Fe 4    & $0.163 \pm 0.012$     & $0.186 \pm 0.005$     & $5.0^{+2.1}_{-1.2}$   & $5.56^{+0.20}_{-0.19}$  & $0.29^{+0.13}_{-0.07}$        & $0.54^{+0.59}_{-0.23}$        & $2500^{+2800}_{-1100}$  \\
\noalign{\smallskip}
Kn 2    & $0.254 \pm 0.003$     & $0.307 \pm 0.008$     &  & $8.40 \pm 0.60$    &         &       &       \\
\noalign{\smallskip}
Kn 40   & $0.276 \pm 0.004$     & $0.269 \pm 0.007$     & $3.5^{+3.9}_{-1.3}$   & $6.4 \pm 0.4$   & $0.13^{+0.15}_{-0.05}$        & $0.32^{+1.07}_{-0.20}$        & $1300^{+4300}_{-\phantom{0}800}$        \\
\noalign{\smallskip}
Kn 45   & $0.122 \pm 0.005$     & $0.151 \pm 0.023$     & $1.8^{+1.1}_{-0.5}$   & $2.95^{+0.28}_{-0.26}$  & $0.032^{+0.020}_{-0.009}$     & $0.27^{+0.44}_{-0.14}$        & $61^{+97}_{-30}$        \\
\noalign{\smallskip}
Pa 12$^{1}$             & $0.553 \pm 0.013$     & $0.602 \pm 0.014$     &       & $5.0^{+0.6}_{-0.5}$     &       &       &       \\
\noalign{\smallskip}
Pa 15$^{2}$     & $0.208 \pm 0.005$     & $0.210 \pm 0.040$     & $3.3^{+1.3}_{-0.8}$   & $\approx10.1$   & $\approx0.130$        & $\approx0.09$ & $\approx1500$         \\
\noalign{\smallskip}
Pa 157$^{1}$    & $0.125 \pm 0.006$     & $0.179 \pm 0.006$     & $3.0^{+0.7}_{-0.5}$   & $5.82 \pm 0.14$ & $0.128^{+0.029}_{-0.020}$     & $0.17^{+0.09}_{-0.05}$        & $1010^{+500}_{-290}$    \\
\noalign{\smallskip}
Pa 26$^{1}$     & $0.143 \pm 0.003$     & $0.05^{+0.07}_{-0.05}$        &       & $13.1^{+1.6}_{-1.4}$    &       &       &       \\
\noalign{\smallskip}
Pa 3$^{1}$      & $1.370 \pm 0.080$     & $0.424 \pm 0.005$     & $0.99^{+0.03}_{-0.03}$         & $0.82 \pm 0.01$       & $0.122 \pm 0.004$     & $0.70 \pm 0.05$       & $25.8^{+1.7}_{-1.6}$    \\
\noalign{\smallskip}
Pa 41   & $0.327 \pm 0.007$     & $0.310 \pm 0.060$     & $2.08^{+0.40}_{-0.29}$        & $5.6^{+0.6}_{-0.5}$     & $0.05 \pm 0.01$       & $0.091^{+0.043}_{-0.027}$     & $880^{+460}_{-290}$     \\
\noalign{\smallskip}
Pre 8$^{1}$   & $0.061 \pm 0.005$       & $0.108 \pm 0.010$     &       & $5.6 \pm 0.6$   &       &       &       \\
\noalign{\smallskip}
\hline
\end{tabular} 
\tablefoot{  
\tablefoottext{1}{Object shows an infrared excess, multicomponent fit performed.} 
\tablefoottext{2}{Parameters uncertain due to star lying near border of theoretical grid;}  
}
\end{table*}

For the calculations of the \Gaia\/ distances, a Gaussian distributed array of parallaxes was generated, 
with the distance being the median of this distribution (see \citealt{Irrgang+2021} for further 
details). The \emph{Gaia} parallaxes were corrected for the zero-point bias using the 
Python code provided by \cite{Lindegren+2021},\footnote{\url{https://gitlab.com/icc-ub/public/gaiadr3_zeropoint}} 
and the parallax uncertainties were corrected by Eq.~16 in \cite{El-Badry+2021}.

In addition, spectroscopic distances can be determined from the SED fits via 
$d_{\mathrm{spec}}= 2 R/\Theta$. In this case, the radius was calculated from the 
Kiel mass and surface gravity via $R=\sqrt{M G/g}$.
In Table~\ref{tab:SEDres}, we also provide an overview of the derived values for the reddening, $E(44-55)$; the spectroscopic distances; and if the 
\emph{Gaia} parallax has a relative error of less than 40\%, the zero-point corrected 
\emph{Gaia} distances. For most objects, the reddening derived in our SED fits agrees fairly well with what is reported by the 2D dust map of \cite{Schlafly2011}, which provides an upper limit on the reddening caused
by interstellar dust. Only for Abell\,16, Pa\,157, and Pre\,8 did we find significantly higher values
for the reddening compared to \cite{Schlafly2011}, suggesting that the PN itself contributes notably to
the reddening.

\section{Discussion}
\label{sec:discussion}

\subsection{Distances}

\begin{figure}
\centering
\includegraphics[width=\hsize]{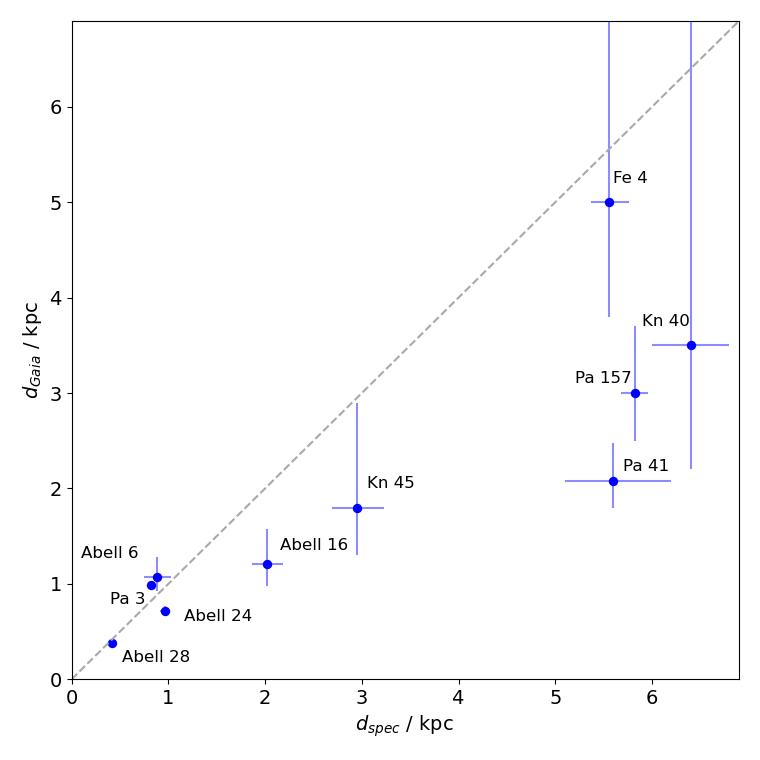}
\caption{
\emph{Gaia} versus spectroscopic distances of our stars. The dashed line indicates a one-to-one correlation.
\label{fig:distances}
}
\end{figure}

By comparing distances derived from the \Gaia\/ parallax with the spectroscopic distances, it is possible to uncover possible systematic errors. In Fig.~\ref{fig:distances}, we plot the \Gaia\/ versus spectroscopic distances. We observed that the \Gaia\/ distances agree within the error limits with the spectroscopic distances only for Abell~6, Fe~4, and Kn~40. For the majority (70\%) of the central stars, the spectroscopic distances significantly exceed the \Gaia\/ distances.

The only object whose spectroscopic distance is actually significantly, but still only slightly, larger than the \Gaia\/ distance is the sdOB star Pa~3. We note that the atmospheric parameters derived for sdOB stars typically do not suffer from large systematic uncertainties. Therefore, the mismatch in the \Gaia\/ distance ($d_{\mathrm{Gaia}}=992^{+30}_{-29}$\,pc) and the spectroscopic distance ($d_{\mathrm{spec}}=$821\,pc) is most likely a result of the assumed canonical mass of 0.48\,\Msol. If one were to assume a mass of 0.70\,\Msol, a perfect agreement with the \Gaia\/ distance would be found ($d_{\mathrm{spec}}(M=0.70$\Msol$)=992$\,pc). 

The fact that spectroscopically derived distances for CSPNe have a tendency to be overestimated when compared to those derived by other methods has been reported in several previous studies \citep{Napiwotzki+2001, Schoenberner+2018, SchoenbernerSteffen2019, Frew+2016}. A solution to this problem is beyond the scope of this paper, but we nevertheless briefly discuss possible reasons for it. 

In principle, an overestimated spectroscopic distance could be caused by an overestimated effective temperature and/or underestimated surface gravity and reddening. We stress that the errors on the spectroscopic distances only assume the formal fitting errors from our $\chi^2$ spectral fits. Yet, as already discussed in \se{sect:analysis}, the atmospheric parameters of very hot stars that are derived by Balmer-line analysis suffer from large systematic uncertainties, and only a UV-metal line analysis could help reduce the systematic errors. On the other hand, again, missing metal-line blanketing and lacking UV spectra might not offer a trivial explanation for this problem. This was shown by \cite{Ziegler+2012} and \cite{Loebling+2020}, who performed a sophisticated UV spectral analysis of the exciting star of \object{Abell 35} and also found that the spectroscopic distance of this star remains too large, and consequently, its gravity mass would be unrealistically small, which is also the case for our sample. We additionally point out that if the observed flux was overestimated for some reason (possible because the PN or a companion adds additional flux), then the discrepancy between the spectroscopic and \Gaia\/ distances would become even larger.

Finally, it is important to note that the spectroscopic distances depend on the assumed mass and on stellar evolutionary computations (e.g., core composition and thickness of the H- or He-envelope). If for some reason the Kiel mass was overestimated, then this could also explain an overestimated spectroscopic distance.

\subsection{Infrared excesses}

Six of our targets were found to show a near- and\slash or far-infrared excess. One of them is the DAO WD Abell~16, which as discussed above was already proposed to show an $I$-band excess due to an M3\,V stellar companion. Our SED fit for this object revealed $\Teff=4200^{+1000}_{-600}$\,K and a surface ratio relative to that of the DAO star of $58^{+3}_{-25}$. This effective temperature would rather indicate a late K-type companion, but within the error limits, an M3\,V-type companion is also possible \citep{Maldonado+2015}. From the surface ratio and the \Gaia\/ parallax, one can deduce a radius of $R=0.14^{+0.06}_{-0.05}$\,\Rsol. This would be about a factor of two to three too small for a mid- or early-type M-dwarf \citep{Maldonado+2015}. However, we point out that our SED analysis could still suffer from additional systematic errors, for instance, because we do not know the metallicity of the M-dwarf. Therefore, we suspect that the factor of two to three agreement of our determined radius as well as what is theoretically predicted is a sign that Abell~16 indeed has a physically connected late-type companion.

A mid-infrared excess was found for the O(H)-type stars Pa\,12, Pa\,26, and Pre\,8 as well as for the sdOB Pa\,3, and it can be reproduced with blackbody effective temperatures between 800\,K and 2690\,K. The derived surface ratios relative to the CSs lie between 140 and 3700. Late M dwarfs or T dwarfs, which have similar \Teff, have radii comparable to that of hot pre-WDs; therefore, an unresolved and physically connected late-type companion seems unlikely to be the (only) source of this excess. A chance alignment of two such special objects (a CS and a foreground late M or even T dwarf) also seems  doubtful. The origin of the mid-infrared excess might thus likely be due to hot dust \citep{Phillips+2005, HolbergMagargal2005}. 

For two of the O(H)-type CSPNe that exhibit a mid-infrared excess---Pa~12 and Pa~26---we additionally found a far-infrared excess. In addition, Pa~157 shows a far-infrared excess only, and for Pre~8, we remain skeptical about the excess found in the W3 band due to the low S/N and the lack of a W4 magnitude. These far-infrared excesses can be reproduced with blackbody temperatures around 100\,K and large surface ratios relative to the CSs, which is typical for cold dust disks \citep{Chu+2011, Clayton+2014}

\subsection{Evolutionary status}

 In the 15 stars with available ZTF light curves, we did not detect any significant periodic photometric variability larger than a few hundredths of a magnitude. This indicates that the majority of stars in our sample represent the hottest phase during the canonical evolution of a single star when transitioning from a AGB star into a WD.

As shown in Figure~\ref{fig:evolution}, nearly all of our objects, except Pa~3, can be considered post-AGB stars. They form a canonical evolutionary sequence from O(H) stars to DAO and DA WDs. Their masses range between 0.45 and $0.67$\,\Msol, with a mean of $0.56\pm0.06$\,\Msol. This agrees well with the mean mass of a sample of hot DAO WDs investigated by \citet{Gianninas2010} with NLTE models: $0.58\pm0.06$\,\Msol.

As for the He/H abundance ratios, for the O(H) stars, they scatter around the expected solar value of 0.1 (from about 0.4 to 1.9 times solar). For the DAO WDs, He is systematically depleted (about 0.2 to 0.4 times solar). This is due to the onset of gravitational settling of helium \citep{UnglaubBues2000}. In the case of our two DA WDs, this process has removed helium from their atmospheres, particularly because of their high surface gravities. This is also demonstrated in Fig.~\ref{fig:evolution}, where several DAO WDs lie to the left of the green dashed line, which indicates where the He abundances should have decreased down to $\log\mathrm{(He/H)} = -3$ according to the predictions of \cite{UnglaubBues2000}. In contrast, the DA WD Abell~28 lies past this line, and our only other DA star, Abell~24, lies very close to it. We note that for such high \Teff---as found for Abell~24 (109.2\,kK)---and low-resolution spectra, the predicted \Ionw{He}{2}{4686} line is very weak, even for $\log\mathrm{(He/H)} = -2$, and higher-resolution spectra might perhaps show that Abell~24 is actually a DAO WD\null. The high effective temperature of this star is also noteworthy, as there appears to be a striking paucity of H-rich WDs relative to their H-deficient counterparts at the very hot (\Teff$>100$\,kK) end of the WD cooling sequence \citep{Werner+2019}. 

Three of our CSPNe, Pa~12, Kn~45, and Abell~6, might be considered as possible candidates for post-RGB CSPNe. Until now, no definitively post-RGB CSPN has been identified, but nine candidates have been suggested in the literature \citep{Jones+2022, Reindl+2023}. For some of them, the post-RGB candidate status is based only on a low (i.e., $M<0.5$\,\Msol) Kiel and sometimes gravity mass, while for other post-RGB candidates, multiband light curves (and partly radial-velocity fitting) have been employed to support this assumption. We note that our candidates are located close to the lowest-mass post-AGB tracks, and as discussed in Sect.~\ref{sect:analysis}, their atmospheric parameters likely still suffer from systematic errors. Unless their low masses are confirmed by UV spectral analysis (which offers a more reliable determination of the atmospheric parameters) and/or the presence of a close companion, we consider these CSs as only weak candidates for post-RGB CSPNe.

The sdOB star Pa~3 is located close to the EHB and He-main sequence (Figure~\ref{fig:evolution}); hence it can be considered as a He-core burning star. These stars are thought to have lost almost their entire envelope due to an interaction with a binary companion during their earlier evolution but to have still managed to ignite He after the envelope was ejected \citep{Heber2016}. However, since the core He burning lasts for about $10^8$\,yrs in these stars, the detection of a PN at this stage is unlikely. As mentioned in Sect.\,\ref{subsect:pa3}, the nebula was listed as a candidate PN in the HASH database with the note that the nebula spectrum is more like that of an \ion{H}{ii} region. This reinforces our suspicion.

It is interesting to note that we found an unusually high gravity mass of $0.70 \pm 0.05$\,\Msol\ for Pa~3, which is significantly higher than the canonical EHB star mass of $\approx\!0.48\,\mathrm{M}_{\odot}$. Recently, \cite{Lei+2023} derived gravity masses for 664 single-lined hot subdwarfs identified in LAMOST\null. They found that while the mass distribution of H-rich sdB and sdOB stars indeed peaks at $0.48\,\mathrm{M}_{\odot}$, 20\% of these stars have gravity masses above $0.6\,\mathrm{M}_{\odot}$. The formation of these relatively massive sdOB stars is, however, a challenge. \cite{Zhang+2017} studied the merger of a He-core WD with a low-mass main-sequence star, but this channel produces intermediate He-rich ($-1.0< \log \rm(He/H) <1.0$) hot subdwarfs, while Pa~3 is He-poor ($\log \rm(He/H)=-1.56$), and most of the hot subdwarfs formed through this merger channel have masses in the range of $0.48-0.50\,\mathrm{M}_{\odot}$, and only a few of them have masses up to $0.52\,\mathrm{M}_{\odot}$.
An alternative channel to produce single H-rich hot subdwarfs involves a main-sequence star that survived the Type Ia supernova (SN\,Ia) explosion of its former massive WD companion and evolved later on into a hot subdwarf. 
This SN Ia channel predicts the production of hot subdwarfs in the mass range $0.35-1.0\,\mathrm{M}_{\odot}$ but also intermediate He-rich surfaces \citep{MengLuo2021}.

The only other object in our sample whose PN is not yet confirmed as a true PN is the large, faint, and round nebula Abell~28. It is listed only as a likely PN in the HASH database and has a major-axis diameter of 330\,arcsec. Assuming an expansion velocity of 20\,km/s and a distance of 384\,pc, a kinematic age of 15\,kyrs can be estimated. Adopting the atmospheric parameters found in our analysis, the cooling time of the central star would already be $\approx 600$\,kyrs, according to the evolutionary calculations from \cite{M3B2016}. Thus, it is more than one order of magnitude higher than the kinematic age of the PN. Similar problems have been reported, for example, for the three H-rich WD CSs inside \object{PN HDW\,4}, \object{PN HaWe\,5} \citep{Napiwotzki1999}, and \object{PNG 026.9+0.04} \citep{Reindl+2023} as well as for the H-poor WD PN nucleus in the Galactic open cluster M37 \citep{Werner+2023}.

\section{Conclusions}
\label{sec:con}

We carried out a spectral and SED analysis of 17 H-rich CSPNe recorded at the HET, which allowed us to increase the number of hot H-rich CSPNe with NLTE atmospheric parameters by $\approx$20\%. We also investigated the ZTF light curves available for our 15 northern objects, and we found that none of them are photometrically variable (with a typical detection limit of a few hundredths of a magnitude).
As reported in several previous studies, in the majority of cases, the spectroscopic distances of our stars exceed the \Gaia\/ distances, which calls for a systematic investigation. Highlights of our results include six objects that show an infrared excess, which could be due to a late-type companion; hot ($\approx 10^3$\,K) and/or cool ($\approx 100$\,K) dust; a rare DA WD (Abell~24) with \Teff possibly in excess of 100\,kK; and an sdOB star (Pa~3) with an unusually high gravity mass. For the latter object, we note that high-resolution spectroscopy would allow for the derivation of metal abundances and a search for chemical peculiarities. For example, if the SN\,Ia channel applies to this star, then after the supernova, the atmosphere of the surviving star may be polluted by the supernova ejecta, meaning that the star could show an enhancement of iron-peak elements \citep{MengLuo2021}. We also encourage high-resolution UV follow-up, which is possible for a handful of our targets. This would enable determination of a reliable \Teff and detailed metal abundances and testing of whether the problem with the too-high spectroscopic distances is resolved. Apart from the search for photometric variability, an additional search for possible (close) companions could be performed by looking for radial-velocity variability. This could help test the possible post-RGB nature of some of our stars and investigate the binary fraction of CSPNe in general. Finally, infrared spectroscopy would improve the characterization of the infrared excess.


\begin{acknowledgements}
We thank the HET queue schedulers and nighttime observers at McDonald Observatory for obtaining the data discussed here.
We thank Matti Dorsch for allowing us to use his model grid for the analysis of the sdOB star Pa~3.
N.R. is supported by the Deutsche Forschungsgemeinschaft (DFG) through grant RE3915/2-1.
The Low-Resolution Spectrograph 2 (LRS2) was developed and funded by The University of Texas at Austin McDonald Observatory and Department of Astronomy, and by The Pennsylvania State University. We thank the Leibniz-Institut f\"ur Astrophysik Potsdam (AIP) and the Institut f\"ur Astrophysik G\"ottingen (IAG) for their contributions to the construction of the integral-field units.

We acknowledge the Texas Advanced Computing Center (TACC) at The University of Texas at Austin for providing high-performance computing, visualization, and storage resources that have contributed to the results reported within this paper.

The Digitized Sky Surveys were produced at the Space Telescope Science Institute under U.S. Government grant NAG W-2166. The images of these surveys are based on photographic data obtained using the Oschin Schmidt Telescope on Palomar Mountain and the UK Schmidt Telescope. The plates were processed into the present compressed digital form with the permission of these institutions. 

This work has made use of data from the European Space Agency (ESA) mission
{\it Gaia\/} (\url{https://www.cosmos.esa.int/gaia}), processed by the {\it Gaia\/}
Data Processing and Analysis Consortium (DPAC,
\url{https://www.cosmos.esa.int/web/gaia/dpac/consortium}). Funding for the DPAC
has been provided by national institutions, in particular the institutions
participating in the {\it Gaia\/} Multilateral Agreement.

This research has made use of the SIMBAD database, operated at CDS, Strasbourg, France.

\end{acknowledgements}




\bibliographystyle{aa}
\bibliography{BB}





\end{document}